\documentclass[
superscriptaddress,
amsmath,
amssymb, 
aps, 
longbibliography,
notitlepage,
12pt,
pre,
floatfix 
]{revtex4-2}

\usepackage{graphicx}
\usepackage{dcolumn}
\usepackage{bm}
\usepackage{hyperref}

\usepackage{amsfonts}       
\usepackage{amsmath}       
\usepackage{amssymb}
\usepackage{nicefrac}
\usepackage{float}
\usepackage[lofdepth,lotdepth]{subfig}
\usepackage{graphicx}
\usepackage{dcolumn}
\usepackage{bm}
\usepackage{hyperref}
\usepackage{xcolor}  
\usepackage{array}
\usepackage[export]{adjustbox}
\usepackage{multirow}
\usepackage[figure,table]{totalcount}
\usepackage{verbatim}
\usepackage{ragged2e}
\usepackage[aboveskip=1pt,labelfont=bf,labelsep=period,singlelinecheck=off]{caption}

\newcommand{\avg}[1]{\left\langle#1\right\rangle} 

\begin{document}

\title{Homeostatic Criticality in Neuronal Networks}

\author{Gustavo  Menesse}
\affiliation{Departamento de Física, FFCLRP, Universidade de São Paulo, Ribeirão Preto, SP, 14040-901, Brazil}

\author{Bóris Marin}
\affiliation{Centro de Matemática Computação e Cognição, Universidade Federal do ABC, 
São Bernardo do Campo, SP, 09606-070, Brazil}

\author{Mauricio Girardi-Schappo} 
\affiliation{Department of Physics, University of Ottawa, Ottawa, ON, K1N 6N5, Canada}

\author{Osame Kinouchi}
\email{okinouchi@gmail.com}
\thanks{Corresponding author}
\affiliation{Departamento de Física, FFCLRP, Universidade de São Paulo, Ribeirão Preto, SP, 14040-901, Brazil}

\begin{abstract}
In self-organized criticality (SOC) models, as well as in standard phase transitions, criticality is only present for vanishing external fields $h \to 0$. Considering that this is rarely the case for natural systems, such a restriction poses a challenge to the explanatory power of these models. Besides that, in models of dissipative systems like earthquakes, forest fires, and neuronal networks,  there is no true critical behavior, as expressed in clean power laws obeying finite-size scaling, but a scenario called “dirty'' criticality or self-organized quasi-criticality (SOqC). Here, we propose simple homeostatic mechanisms which promote self-organization of coupling strengths, gains, and firing thresholds in neuronal networks. We show that with an adequate separation of the timescales for the coupling strength and firing threshold dynamics, near criticality (SOqC) can be reached and sustained even in the presence of significant external input.
The firing thresholds adapt to and cancel the inputs ($h$ decreases towards zero). 
Similar mechanisms can be proposed for the couplings and local thresholds in spin systems and cellular automata, which could lead to applications in earthquake, forest fire, stellar flare, voting, and epidemic modeling.
\end{abstract}


\maketitle

\section*{Highlights}
\begin{itemize}
    \item We introduce a novel mechanism that promotes Self-organized quasicriticality (SOqC) even in the presence of external fields;
    \item Criticality is asymptotically approached as the ratio between the homeostatic time scales increases;
    \item The mechanism is general and can be applied to any phase transition. 
\end{itemize}

\section{Introduction}

The idea of self-organized criticality (SOC)~\cite{Bak1987}, in which
a given dynamical system has a critical point as an attractor, without  
any ad hoc imposition of parameter values (fine-tuning), 
in some sense has never truly been achieved. 
The most successful models under this ideal display bulk conservation, 
such as in the Abelian sandpile~\cite{Jensen1998,Dickman1998,Dickman2000}. 
However, this conservation requirement can also be seen as a form of fine-tuning,
since the number of dissipated grains in the spread of avalanches must be zero.
Moreover, the infinite separation of time scales between driving and
avalanches in SOC models can be viewed as yet another form of fine-tuning.

When we consider dissipative systems such as earthquakes, forest fires or
neural networks, we find that only self-organized quasicriticality (SOqC)
holds~\cite{Bonachela2009,Buendia2020}. 
This regime is characterized by the system performing stochastic oscillations
around the critical point.
Several of such models include continuous drive and dissipation, some 
of which can be viewed as homeostatic mechanisms that drive the network 
toward the critical point.

In the case of cortical models, SOqC mechanisms have been widely employed 
to explain the experimental observation of neuronal
avalanches~\cite{Beggs2003,Beggs2008,Chialvo2010,Munoz2018,Carvalho2021}. 
The main studied homeostatic mechanisms are related to synaptic
dynamics~\cite{Levina2007,Bonachela2010,Zeraati2021},
but dynamical gains~\cite{Brochini2016,Costa2017,Campos2017,Kinouchi2019}
and firing thresholds have also been
considered~\cite{delPapa2017,Girardi2020,Girardi2021}
(for a review see~\cite{Kinouchi2020}). 

In the absence of homeostatic mechanisms, a critical regime is obtained only 
with strong and non-local fine-tuning over, for example, all coupling weights 
(synapses) $W_{ij}$, so that the distribution $P(W_{ij})$ must have average
$\avg{W_{ij}} \equiv W = W_c$ (the control parameter).
With homeostasis, this constraint is relaxed: now we can start from any
distribution $P_{t=0}(W_{ij})$ and, 
after a transient (the self-organization process), one obtains a stationary
$P^*(W_{ij}) \equiv \lim_{t\rightarrow \infty}P(W_{ij})$
where 
$W^* \equiv \lim_{t\rightarrow \infty}\avg{W_{ij}} \approx W_c$. 
Similar reasoning applies to neuronal gains $\Gamma_i$ and firing thresholds $\theta_i$. 

One important aspect in any SOC model is that phase transitions, 
and therefore criticality, exist only for zero or very small external field~\cite{Dickman2000,Williams-Garcia2014,Decandia2021},
so any homeostatic mechanism will need to self-organize the system 
to a state where the effective external field vanishes. 

Here, we propose a solution to this problem by unveiling an interplay 
between homeostatic time scales, network size and external field.
We show that the important factor behind the generation of power-law 
avalanches that obey the size-duration scaling law is sensory 
adaptation, mediated by dynamic firing thresholds.
This adaptation must occur over long periods of time when compared 
to the intrinsic neuronal and synaptic time scales.

We develop a mean-field theory for the homeostatic model, and compare 
it to simulation results for a sparse random network with $K$ input
neighbors per node.
These mechanisms are simple and very general: they can be adapted 
to systems composed of other units such as spins, cellular automata, 
discrete time maps or continuous time neuronal models with
pulse coupling given by weights $W_{ij}$. 

\section{The model and its mean-field approximation}

We consider a network of $N$ discrete-time stochastic leaky integrate-and-fire
neurons~\citep{Gerstner1992,Galves2013,Larremore2014,Brochini2016,Costa2017,Kinouchi2019,Zierenberg2020}. 
A binary indicator $X_i \in \{0,1\}, i = 1,\ldots,N$,
denotes silence ($X_i = 0$) or the firing of an action potential (spike, $X_i = 1$). 
The membrane potential of neuron $i$ evolves according to:
\begin{equation}\label{eq:mod2D}
    V_{i}(t + 1) = \mu_i V_{i}(t) + I_i + \frac{1}{K}
    \sum_{j=1}^{N}  W_{ij} X_j(t)\:,
\end{equation}
where $0 \leq \mu_i \leq 1$ are leakage parameters and 
$I_i$ are external inputs.
The directed synaptic weight matrix $W_{ij}$ has exactly 
$K$ incoming links from $j$ to $i$. The outgoing links, by this
construction, have a binomial distribution with average $K$  
and standard deviation 
$\sigma  = \sqrt{ K(1-K/(N-1))}$. Notice that while the sum is over
all $N$ potential neighbors (all $W_{ij}$ entries, most of
them are zero for finite $K$),   
the normalization considers only the true $K$
(not $N$) neighbors. In the case of a complete
graph, we use $K = N-1$ given that $W_{ii}= 0$.

If a neuron fires at time step $t$, its membrane 
potential is reset, $V_i(t + 1) = 0$. Otherwise, its voltage is updated according to
Eq.~\eqref{eq:mod2D}. A spike occurs with probability
\begin{equation}
    \label{eq:probfire}
    P\left( X_i(t) = 1 \:|\: V_i(t)\right) \equiv \Phi(V_i(t))\:,
\end{equation}
where $\Phi(V)$ is the so-called firing function. 
The model incorporates an
absolute refractory period of one time step by imposing 
$\Phi(0)= 0$.
For this class of models, there are no strong requirements on the firing function $\Phi$ besides a sigmoidal shape.
For analytical convenience, we adopt a
linear-saturating shape~\cite{Larremore2014,Brochini2016,Girardi2020,Girardi2021}:
\begin{equation}
\label{eq:phi}
\Phi(V_{i})=
\begin{cases}
    0     & \text{if } V_{i} < \theta_i \\
    
    \Gamma_{i}\left( V_{i}-\theta_{i}  \right) & \text{if }  \theta_i < V_{i} < V_i^S \\
    
    1 & \text{if } V_i > V_i^S 
    \end{cases} \end{equation}
where $V_i^S=1/\Gamma_{i}+\theta_{i}$ is the
saturation potential. Here, $\theta_i$ represents a firing threshold for neuron
$i$. This choice of firing function implies that
there is a finite probability of the $i$-th
neuron emitting a spike only when $V_i(t)>\theta_i$, which 
increases linearly with $V_i$.

In the absence of homeostatic tuning (which we call the static model), assuming that the distribution $P(W_{ij})$ has finite variance,
the average synaptic weight $W \equiv \avg{W_{ij}}$ can be taken as a control parameter.  
The same applies to the neuronal gains $\Gamma_i$,
firing thresholds $\theta_i$, leakage parameters $\mu_i$ and inputs $I_i$, so that $\Gamma = \avg{\Gamma_i}$ and $\mu = \avg{\mu_i}$
can also be considered as control parameters. Interpreting $\theta = \avg{\theta_i}$ as the average local field (local adaptation current) and $I = \avg{I_i}$ as the average external field (external input current), we have that $h = I - \theta$ is the total or effective field.

The fraction of spiking neurons, also known as firing density or firing rate,
$ \rho(t) = \avg{X_i(t)} \equiv \frac{1}{N}\sum_{i=1}^N X_i(t)$ 
represents the activity of the network. The time average of $\rho(t)$ is
calculated in the steady state and is the relevant order parameter.
 A summary of variables and parameters used in this work is presented in Table~\ref{tab:table1}.
 
 \begin{table}[ht!]
\vspace{1.5em}
{\scriptsize
\caption{\label{tab:table1} Variables and parameters used in the model.}
\begin{tabular}{cc}
\toprule 
\textbf{Variable} &  \textbf{Symbol} \\\hline
 Neuron state (binary) & $X_{i}$  \\ 
 Membrane potential & $ V_{i} $  \\
 Saturation potential & ${V_i}^{S}$ \\
 External field (input current) & $I_i$\\
 Firing probability function & $\Phi$ \\
 Firing density (activity)& $\rho$ \\
Effective external field & $h_i$  \\
 Neuronal gain & $\Gamma_i$  \\
 Synaptic weight & $W_{ij}$  \\
 Firing threshold & $\theta_i$\\\hline
\end{tabular}
\begin{tabular}{cc}
\toprule 
\textbf{Parameter} &  \textbf{Symbol}  \\ \hline
Number of input neighbors & $K$ \\
Leakage parameter & $\mu_i$\\
Neuronal gain recovery time & $\tau_\Gamma$ \\
Synaptic weight recovery time & $\tau_W$ \\
Neuronal gain depression & $U_\Gamma$  \\
Synaptic weight depression & $U_W$  \\
Synaptic weight baseline & $A_i$\\
Neuronal gain baseline   & $B_i$\\
Time scale ratio: threshold/synaptic & $a$ \\
Depression ratio: threshold/synaptic & $b$   \\\hline
\end{tabular}
}
\vspace{1.5em}
\end{table}

When $\Gamma$, $W$ and $\theta$ are fixed (static model), a mean-field approximation (equivalent to taking the $K\to\infty$ limit) can be calculated from:
\begin{equation} \label{map}
    \rho(t+1) = \int \Phi(V) P_t(V) \:dV \:,
\end{equation}
where $P_t(V)$ is the distribution of voltages at time $t$ ~\citep{Brochini2016,Kinouchi2019}.

For $\mu = 0$, considering the case where the stationary potentials fall within the linear ($0<V_i<V_i^{S}$) branch of equation \eqref{eq:phi}, the solution leads to the mean-field map:
\begin{equation}\label{rhot}
\rho(t+1) = (1-\rho(t))\Gamma (W\rho(t)+h)\:\:.
\end{equation}
The steady state is the fixed point of equation \eqref{rhot}:
\begin{equation}
    \rho^\pm= \frac{\Gamma W -1 - \Gamma h \pm 
    \sqrt{(\Gamma W -1 - \Gamma h)^2 + 4 \Gamma^2 W  h }}{2 \Gamma W} \:.
\end{equation}
The phase transitions of this model are given by the bifurcations of
the map in equation \eqref{rhot}~\cite{Girardi2019}.
When the field $h$ is negative, we have a discontinuous 
(first order) phase transition and, when $h > 0$, there is always activity $\rho > 0$ and no transition~\citep{Brochini2016,Girardi2020}.

\begin{figure}[htp]
\begin{center}
	\includegraphics[width=\textwidth]{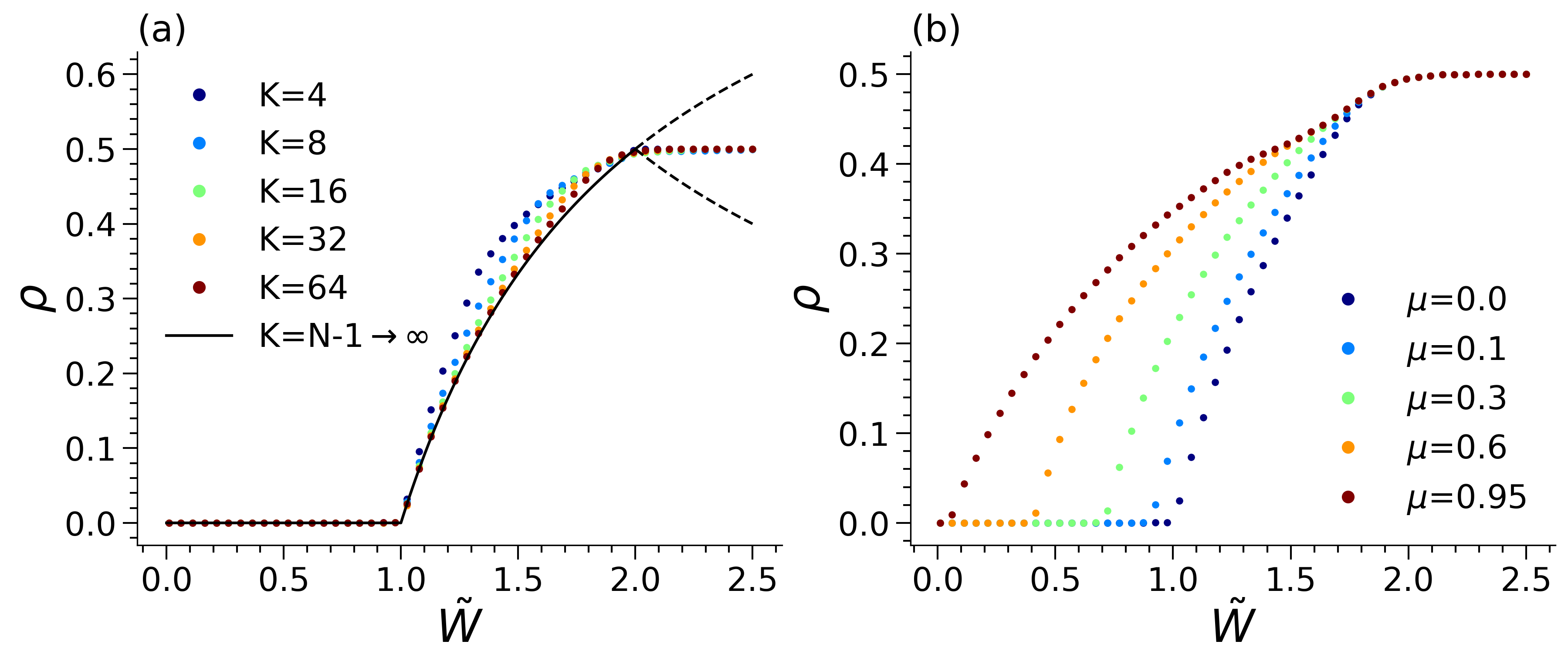}
\caption{a) Order parameter $\rho(\tilde{W};\mu = 0$) as a function of
$\tilde{W}$ for different number of input neighbors $K$. From left to
right, $K = 4, 8, 16, 32$ and mean-field (solid). The bifurcation at
$\tilde{W}=2$ leads to the creation of a period$-2$ 
synchronous regular state~\cite{Brochini2016,Girardi2020}.
b) $\rho(\tilde{W};K=32)$ for various leakage parameters
$\mu$. From right to left, $\mu = 0, 0.1, 0.3, 0.6$ and
$0.95$. Network size $N = 10^4$. The leakage parameter does not
change the critical exponents, only shifts the critical point~\cite{Girardi2021}.
}
\label{fig1} 
\end{center}
\end{figure}

For $h=0$, we have a second order phase transition, which is given by:
\begin{equation}
    \rho(W,\Gamma) \sim \left(\frac{ W - W_c(\Gamma)}{ W_c(\Gamma)} \right)^{\beta}
\end{equation}
for $W > W_c= 1/\Gamma$, while $\rho = 0$ (absorbing state) for 
$ W < W_c$. The order parameter exponent is $\beta = 1$.
The hyperbola $W_c(\Gamma) = 1/\Gamma$ is a critical line in
the $W \times \Gamma$ plane. Similarly to what occurs in the Ising model,
where the important variable is the combined quantity $J/T$, here the
important variable is $ \tilde{W} \equiv \Gamma W$, which defines the
critical point $\tilde{W}_c =  1$ (see Fig. \ref{fig1}).
At $\tilde{W}=2$, a period-2 orbit is created. However, it 
is not relevant to the present work and has been studied in detail 
elsewhere~\citep{Brochini2016,Girardi2020}.

For random networks, a continuous phase transition is also observed:
\begin{equation}
    \label{eq:rhocrit}
    \rho(\tilde{W};\mu) = C(K,\mu) \left( \frac{ \tilde{W} - \tilde{W}_c(\mu)}{\tilde{W}}\right) \:,
\end{equation}
which can clearly be seen in Fig. \ref{fig1}. 
The critical point is independent of $K$ (see the Supplementary Material 
for details), but depends on $\mu$ according to 
$\tilde{W}_c(\mu) = (1-\mu) \tilde{W}_c(0)$ 
similarly to the infinite $K$ limit~\cite{Girardi2021}. 
Meanwhile, $\beta = 1$ is independent of $K$ and $\mu$, which is
compatible with the finding of mean-field directed percolation (DP)
exponents in a large set of experiments~\cite{Beggs2003,Beggs2004,Carvalho2021}. 
Since the critical exponents of many branching processes-like universality classes
coincide at the mean-field level (\textit{e.g.} DP,  Manna
~\cite{Henkel2008}), we cannot ensure that our model belongs to the DP class.


In order to tune the network to the critical
point~\cite{Levina2007,Bonachela2009,Bonachela2010,Brochini2016,Costa2017,Kinouchi2019,Girardi2020,Kinouchi2020}, we will introduce homeostatic mechanisms for parameters $\Gamma_i$, $\theta_i$ and $W_{ij}$. 
The calculations are done in the mean-field level for $\mu = 0$, but similar results can be shown in
simulations for general $K$ and $\mu$.

First, we impose depressing-recovering dynamics to the control parameter $\tilde{W}(t) \equiv \avg{\Gamma_i W_{ij}(t)}$. Following biological motivations, we propose two mechanisms: 
one for neuronal gains $\Gamma_i(t)$ and another for synaptic weights $W_{ij}(t)$. We use dynamics similar
to the Levina-Hermann-Geisel mechanism~\cite{Levina2007} for each variable:
\begin{eqnarray}
\label{HomDynW}
   W_{ij}(t+1) &=&  W_{ij}(t) + \frac{1}{\tau_W}\left(\frac{A_i(1-\mu_{i})}{\Gamma_i(t)}-W_{ij}(t)\right) \nonumber\\
     &-& U_W W_{ij}(t) X_j(t) \:,\\
\Gamma_i(t+1) & =& \Gamma_i(t) + \frac{1}{\tau_\Gamma}\left(B_i-\Gamma_{i}(t)\right) 
     - U_\Gamma \Gamma_{i}(t) X_i(t) \:. \label{HomDynG}
     \end{eqnarray}
The dynamics for synaptic weights ($W_{ij}$) has a basal level $A_i(1-\mu_{i})/\Gamma_{i}(t)$, 
a recovery time $\tau_W$ and a depressing factor $0< U_W< 1$ related to the fraction of depleted
neurotransmitter vesicles in the synapse due to a presynaptic spike $X_j=1$.
A similar idea applies to the dynamics of membrane excitability (neuronal gains $\Gamma_i$).

The coupling between $W_{ij}(t)$ and $\Gamma_i(t)$ is necessary to get 
$W^* = (1-\mu)/\Gamma^*$, resulting in $\tilde{W}_c= 1-\mu$. 
This is a small non-locality in the basal level of synaptic weights, which introduces a dependence of the
effective recovery time of synapses $\tau_W$ on the neuronal gain $\Gamma_i$
and on the leakage parameter $\mu_i$. 
In biological neurons, this coupling between synapses and neuronal excitability could be mediated by retrograde signals (\textit{e.g.}, active dendritic spikes~\cite{Spruston1995,Gollo2009}).

The $\Gamma_i(t)$ dynamics depends on the local activity $X_i$, referring to the cell body with gain $\Gamma_i$.
Averaging over sites (in the $\mu=0$ case) and neglecting cross-correlations,
the MF equations become:
\begin{eqnarray}
\label{WMF}
     W(t+1) & = &   W(t) + \frac{1}{\tau_W}\left(\frac{A}{\Gamma}-W(t)\right) 
     - U_W W(t) \rho(t) \:,\\
     \label{GMF}
    \Gamma(t+1)  & = & \Gamma(t) + \frac{1}{\tau_\Gamma}\left(B-\Gamma(t)\right) 
     - U_\Gamma \Gamma(t) \rho(t)\:.
\end{eqnarray}

To achieve criticality, we also need $h$ to be 0. For spin systems, 
zero external magnetic field is a natural condition, 
despite being a fine-tuning operation seldom 
discussed in the literature of neuronal
avalanches~\citep{Williams-Garcia2014,Girardi2020,Decandia2021}. 
Here, for integrate-and-fire neurons, this condition is not so natural: we must 
fine-tune $\theta_c = I/(1-\mu)$ in order to achieve $h_c=0$~\cite{Girardi2021}.  
Therefore, we also need a homeostatic mechanism to drive $h$ toward zero.

We propose a simple firing-threshold adaptation mechanism:
\begin{equation}
\label{thetat}
    \theta_i(t+1) = \theta_i(t) - \frac{1}{a\tau_W}\theta_{i} (t)
     + bU_W \theta_{i}(t) X_i(t) \:,
\end{equation}
yielding the average over neurons $\theta(t) \equiv \avg{\theta_i(t)} $
\begin{equation}
     \label{thetaMF}
   \theta(t+1)  =  \theta(t) -\frac{1}{a\tau_W}\theta(t)  \:
    + bU_W \theta(t) \rho(t)\:.
\end{equation} 
Here, the $\theta$ time constant and increment parameters are given as multiples $a$ and $b$, respectively,
of the time constant and depression parameters of the synaptic weight dynamics ($\tau_W$ and $U_W$).

From the 4-dimensional mean-field map, given by the equations~\eqref{rhot}, \eqref{WMF}, 
\eqref{GMF}, and \eqref{thetaMF}, we get the following relevant fixed point:
\begin{eqnarray}
\label{rhoFP}
\rho^* &=& \frac{1}{ab\:\tau_W U_W } \:,\\
\Gamma^* &=& \frac{B}{1+\frac{\tau_\Gamma U_\Gamma}{ab\:\tau_W U_W}} \:,\\
W^* &=& \frac{A}{\Gamma^*(1+\frac{1}{ab})}  \:,\\
\label{sthetafp}
h^{*} &=&I-\theta^{*} = \rho^{*}\left(W^{*}-\frac{1}{\Gamma^{*}}\right)+\frac{\rho^{*2}}{\Gamma^{*}}-\mathcal{O}(\rho^{3})
\end{eqnarray}
Comparing this steady state to the critical point (which requires $\rho_c=0^{+}$,$\tilde{W}_c= W\Gamma=1$, and $h_c=0$), we can see that 
two conditions are needed to reach quasi-criticality. 
First, $ab \gg 1$, expressing a large separation between the $W$ and $\theta$ recovery times, 
which is a very common feature in SOC models~\cite{Pruessner2012}. 
And secondly, we need to fine-tune $A=\avg{A_i}\approx 1$ to obtain $h=\mathcal{O}({\rho^{*}}^{2})\approx 0$ \cite{Girardi2021}.
Notice that under these conditions, the first term in equation~\eqref{sthetafp} disappears,
leaving the correct dependence $h\sim\rho^{*2}$ that, in the static system, defines the field
exponent $\delta_h=2$ for the mean-field DP-like phase transition~\cite{Henkel2008,Dickman1999}.

\section{Simulation results}

For the simulations, we chose $\tilde{W}$ time scales in the order of 10$^2$ ms ($\tau_W=300$ and
$\tau_\Gamma=100$). 
Therefore, $\tilde{W}$ evolves more closely with the network activity propagation dynamics. 
On the other hand, we model the adaptive threshold mechanism as a long-term homeostatic
regulation 
($a> 10^{3}$), which means that this adaptation process occurs on a timescale slower than
that of network dynamics. Recent results show that the mossy cells in the dentate gyrus
present such long recovery threshold time scales~\cite{Trinh2022}.

\begin{figure}[htp]
\begin{center}
    \includegraphics[width=0.8\textwidth]{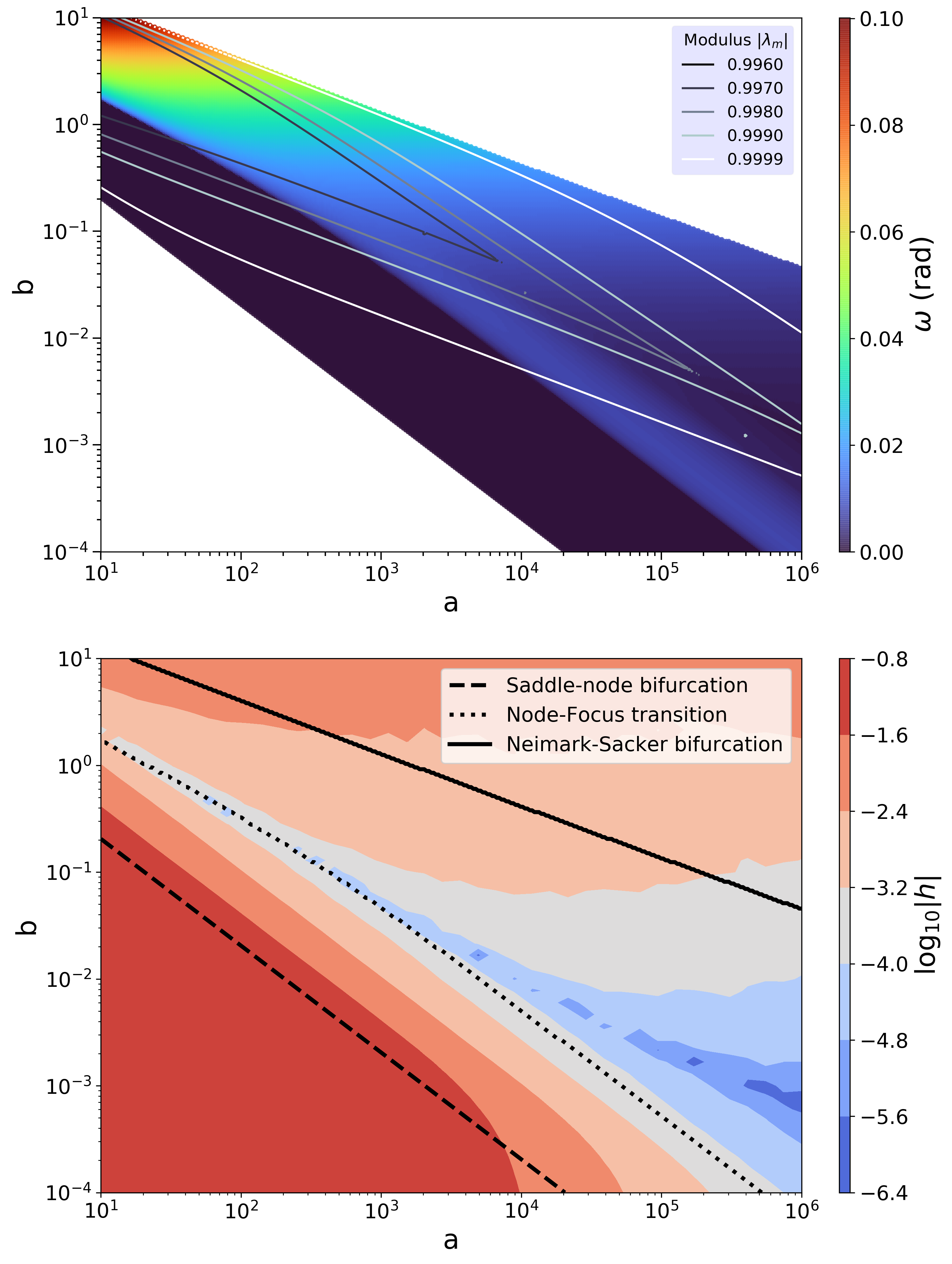}
\caption{\textbf{Mean-field stability diagram}. (top) Argument (heat-map) and modulus (contour lines) of the leading eigenvalue of the 4-dimensional mean-field map $[\rho(t),\Gamma(t),W(t),\theta(t)]$.
Colored regions correspond to systems with a stable fixed point, and white regions to dynamics
with unstable fixed points.
(bottom) Effective field $|h|$ obtained from random network simulations.
Limited power law avalanches,
sufficient to explain experiments, are observed in the blue region. Bifurcation and transition lines are also shown. 
}
\label{fig2} 
\end{center}
\end{figure}

For a robust quasi-critical regime with nearly critical avalanches, the system needs to
evolve towards a stable fixed point not far from the true critical point as quickly as possible.
This should happen while keeping oscillations around the equilibrium to a minimum.
This can be achieved by minimizing the spectral radius of the Jacobian matrix. In Fig. 
\ref{fig2}(top), we study the stability of fixed point ($\rho^*, \Gamma^*, W^*, \theta^*$) with 
respect to time scale separation parameters a,b when A=1. Colored regions indicate dynamics with 
stable fixed points.

Parameter regions with leading eigenvalue modulus $|\lambda_m| < 0.9999$ and argument
$\omega < 0.01$ (top panel of figure~\ref{fig2}) give rise to dynamics with $|h| < 10^{-4}$
(bottom panel of figure~\ref{fig2}). Thus, for a given choice of parameter $a$, there is a range 
of $b$ values (coarse tuning) that allow the homeostatic mechanism to reach and maintain the
$h$ and $\tilde{W}$ values close enough to their critical values [$|h| \sim \mathcal{O}(10^{-4})$
and $\tilde{W}\sim\tilde{W}_c-\mathcal{O}(10^{-2})$], yielding quasi-critical
power-law avalanches that scale with system size for large $a$
(to be shown further ahead in this section). 

\begin{figure}[htp]
\begin{center}
	\includegraphics[width=0.7\textwidth]{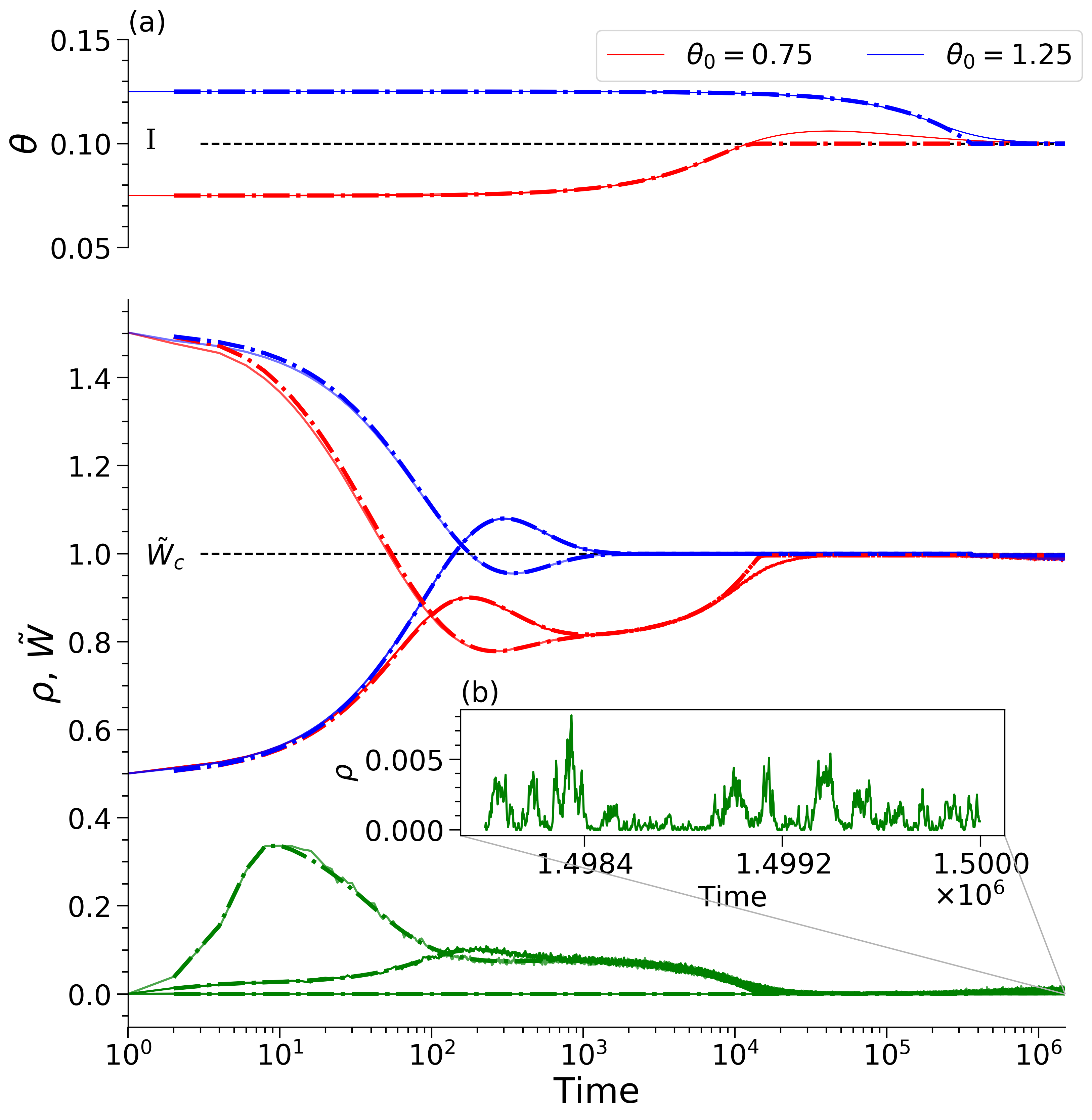}
\caption{{\bf Self-organization of $\tilde{W}(t)$ and $\theta(t)$ from different initial conditions (blue and red).} The target values are $\theta_c = I = 0.1$ (or $h = I-\theta = 0)$ and $\tilde{W}_c = 1$.  (a) Time series for
$\theta(t)$ (top), $\tilde{W}(t)$ (middle) and $\rho(t)$ (green, bottom).  Mean-field (dot-dash lines) and random network ($K=32$, solid lines) simulations with $N=10,000$ neurons.  
(b) Avalanche behavior for stationary $\rho(t)$.
Parameters: $\tau_W=300$,$\tau_\Gamma=100$, $U_W=0.01$, $U_\Gamma=0.01$,$B=1$, $A=1$, $a=5000$ and $b=0.05$.}
\label{fig3} 
\end{center}
\end{figure}

\begin{figure}[htp]
\begin{center}
    \includegraphics[width=0.8\textwidth]{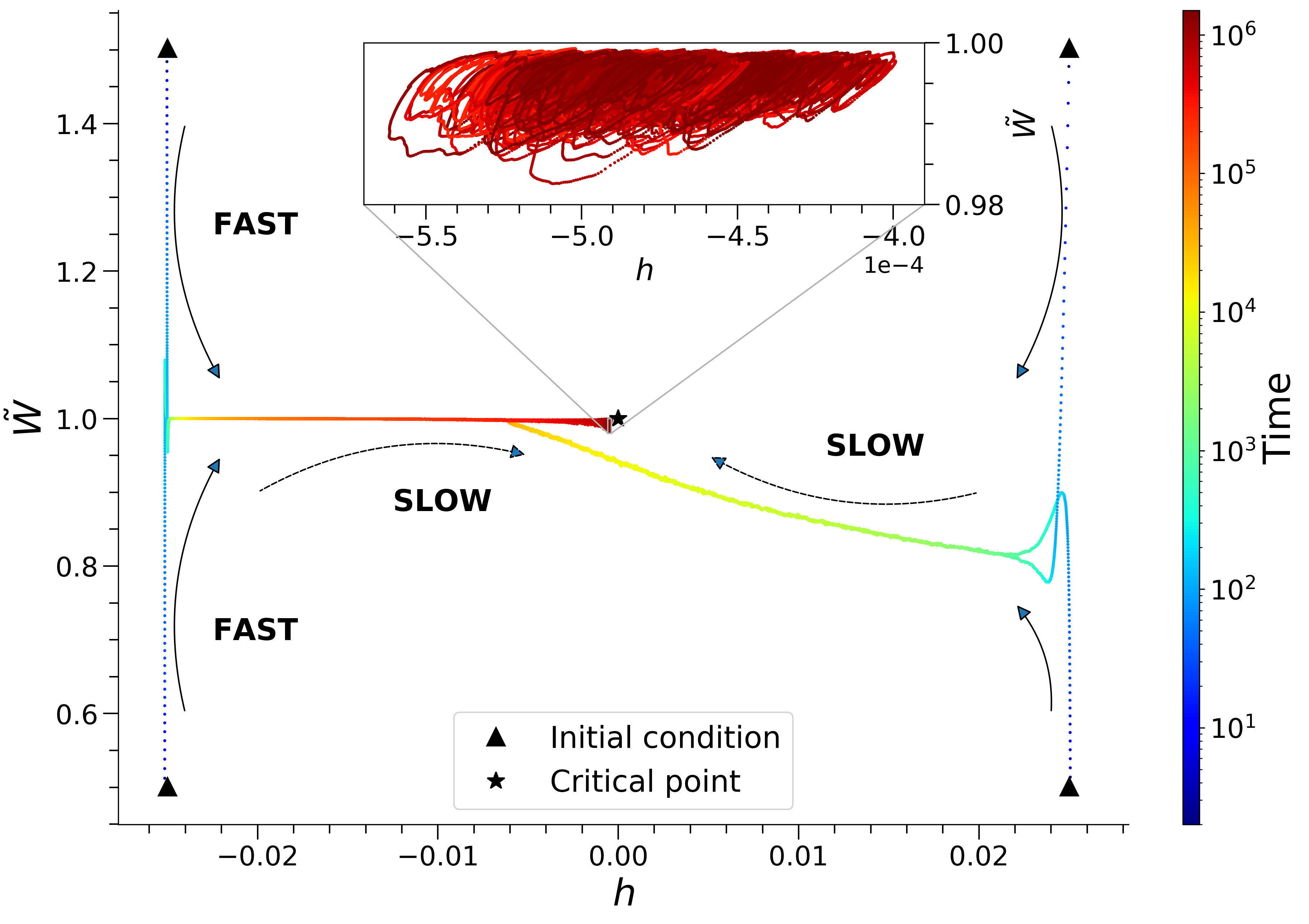}
\caption{\textbf{State space}. Self-organization in the $\tilde{W}\times h$ plane for four different initial conditions (black triangles), input $I=0.1$, in a random network with $K=32$ and $N=10,000$ neurons.
The temporal evolution of the system is indicated by the arrows and the color code
(bluish hues correspond to the vicinity of initial conditions, while reddish to the steady state).
Parameters: $\tau_W=300$, $\tau_\Gamma=100$, $U_W=0.01$, $U_\Gamma=0.01$, $a=5000$ and $b=0.05$.
}
\label{fig4} 
\end{center}
\end{figure}

Results for the mean-field and random network simulations are shown 
in figure~\ref{fig3}a. Initial conditions were chosen from four different sets  
(for a given simulation, $\Gamma_i(0)$ is either 0.5 or 1.5, while 
$\theta_i(0)$ are drawn from normal distributions with mean 0.75 or 1.25 and standard deviation 0.01, and $W_{ij}(0)$ from the uniform 
distribution in $[0;2]$). In all cases, trajectories in the $\tilde{W}\times h$ 
space (see figure~\ref{fig4}) show low amplitude stochastic oscillations around
a slightly subcritical point, with mean amplitude of approximately $0.01$ in 
$\tilde{W}$ and $10^{-4}$ in $h$. In figure~\ref{fig3}b, we show the stochastic
oscillations in the firing rate $\rho(t)$.

We measured the size and duration of $10^{6}$ avalanche events after
disregarding the transient activity. 
Avalanches are defined here as all activity between two consecutive
visits to the absorbing state of the underlying static system ($\rho=0$)~\cite{Dickman1999,Girardi2018}.
In other words, we sum all spikes $N\rho(t)$ from all the time steps between two subsequent instants in which $\rho(t)$
is zero
(\textit{i.e.}, all the activity in each peak between zeros of $\rho(t)$ in figure~\ref{fig3}b).
This method does not require thresholding the activity, as it makes use of the actual
silent state of the network, thereby avoiding known biases in the
estimation of avalanche exponents~\cite{Touboul2017,DiSanto2018,Villegas2019}.

Near the critical point, we expect the avalanche sizes $s$ and duration $d$ to be distributed
according to 
$F(s)= P(S > s) \sim s^{1-\tau}$ and 
$F(d)= P(D > d) \sim d^{1-\tau_d}$ respectively, with exponents $\tau=3/2$ and $\tau_d=2$
(exponents for a mean-field DP-like
branching process~\cite{Henkel2008,Pruessner2012}). 
The $F(x)$ functions are the complementary cumulative distributions,
which are defined as the integral of the distribution $P(X)$ over $X>x$. These distributions
are a convenient choice since they are continuous functions that can be directly calculated from the data and do not rely on binning~\cite{Girardi2013b}.

Having the correct exponents in the avalanche distributions is not a
sufficient condition for identifying criticality \citep{Touboul2017,Girardi2021b}.   
Therefore, we investigated the scaling law between mean avalanche size
and duration $\avg{s}(d)$ leading to the scaling relation
$m_{\text{theory}}=(\tau_d-1)/(\tau-1)$. At the underlying critical point 
(towards which the system is evolving), we have $m_{\text{theory}}=2$. 
In order to compare simulation results to the theory, we define the distance to
criticality coefficient $dcc$ by:
$dcc\equiv|m_{\text{theory}}-m_{\text{fitted}}|$, where the $m_{\text{fitted}}$
is given by directly fitting the data of $\avg{s}$ \emph{versus} $d$.
To compute the $dcc$, we take the mean value of $m_{\text{fitted}}$ adjusted to the
simulation results with different $N$ for three values of $a$ and $b$.

\begin{figure*}[htp!]
\begin{center}
    \includegraphics[width=\textwidth]{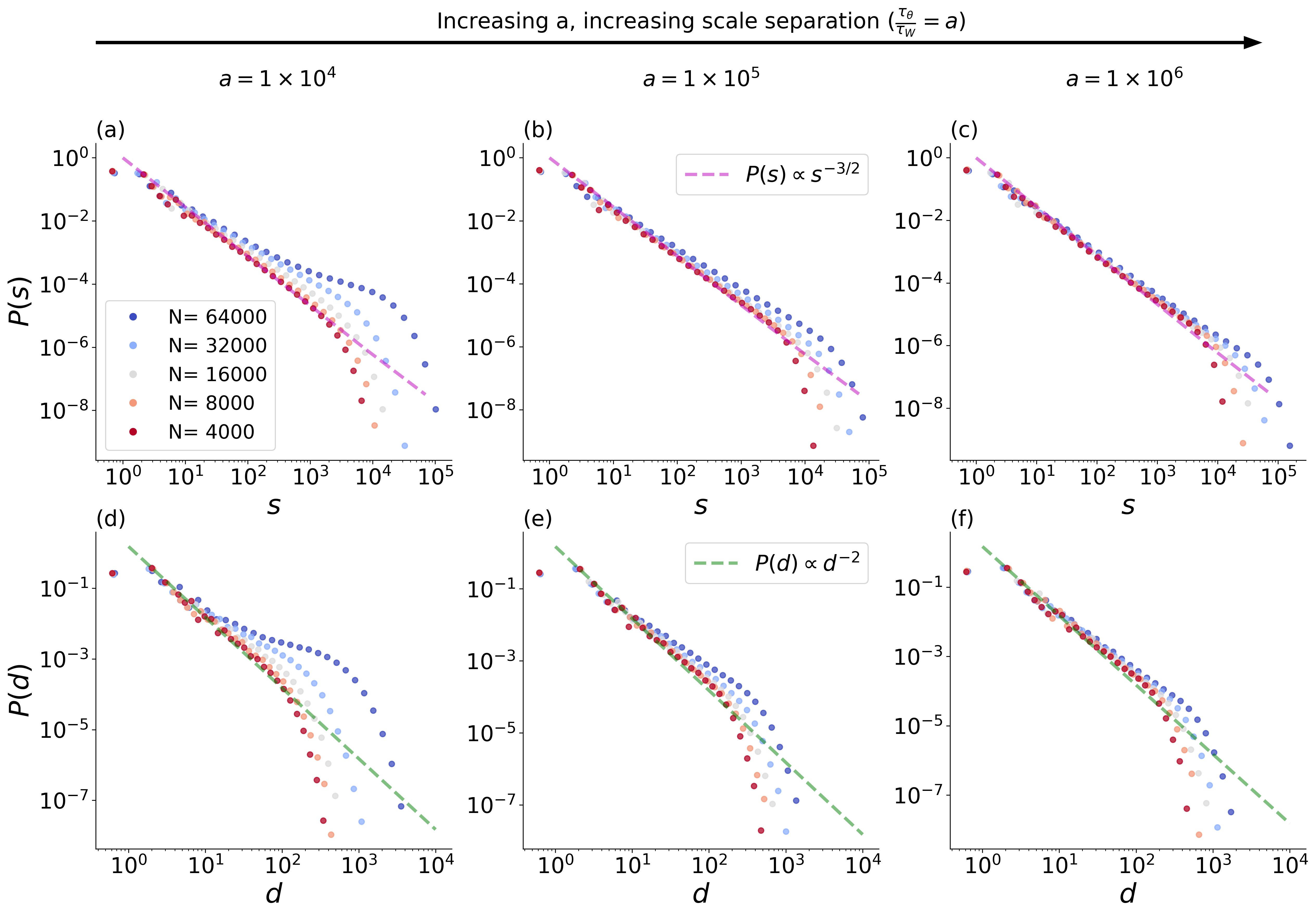}
\caption{\textbf{Power laws in avalanche sizes and duration distributions for increasing values of $a$ and network size $N$}. The first and second rows show the distributions of sizes and duration, respectively. The agreement with the expected power laws (dashed lines) increases with $a$. Results obtained for a quenched simulation of a directed random network with $K=32$. Parameters: (a,d) $a= 10^{4}$ and $b=8\times 10^{-2}$, (b,e) $a= 10^{5}$ and $b= 10^{-2}$ and (c,f) $a= 10^{6}$ and $b= 10^{-3}$.}
\label{fig5} 
\end{center}
\end{figure*}

\begin{figure*}[htp]
\begin{center}
    \includegraphics[width=\textwidth]{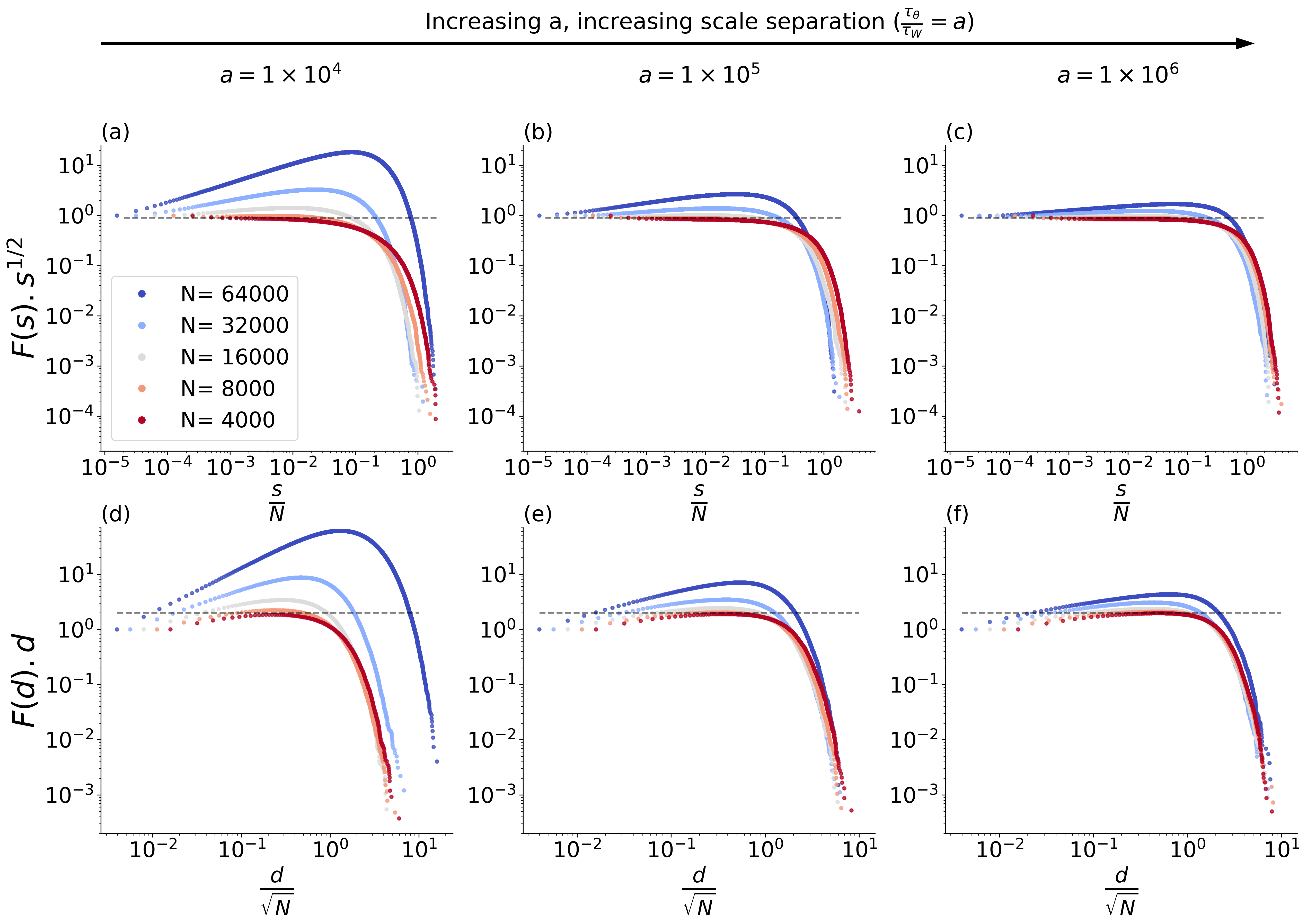}
\caption{\textbf{Finite-size scaling collapse of avalanche size and duration complementary cumulative distributions for increasing values of $a$ and network size $N$}. The first row shows the collapse of the complementary cumulative distribution $F(s)$, and second row for duration, $F(d)$. The collapse improves with increasing $a$. Results obtained for a quenched simulation of a directed random network with $K=32$. Parameters: (a,d) $a= 10^{4}$ and $b=8\times 10^{-2}$, (b,e) $a= 10^{5}$ and $b= 10^{-2}$ and (c,f) $a= 10^{6}$ and $b= 10^{-3}$.}
\label{fig6} 
\end{center}
\end{figure*}

\begin{figure*}[htp]
\begin{center}
    \includegraphics[width=1.\textwidth]{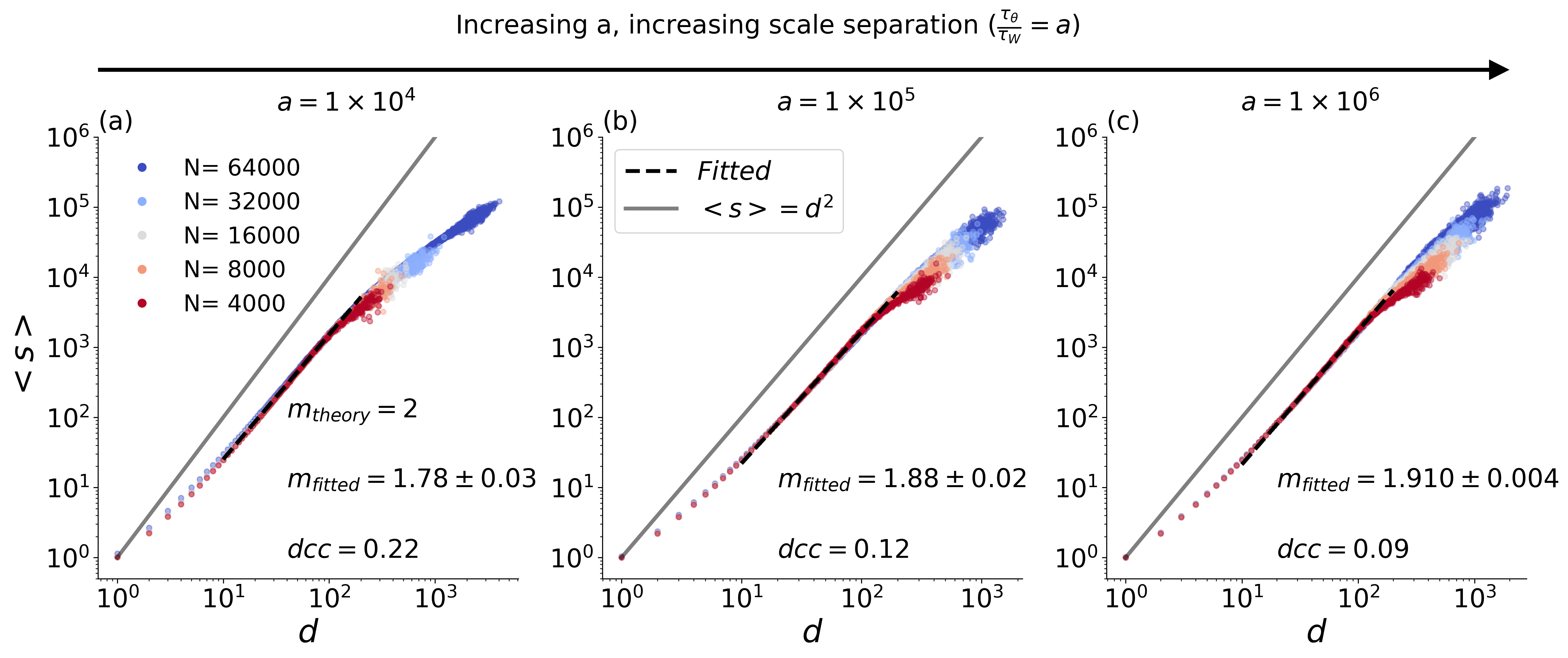}
\caption{\textbf{Average avalanche size vs. duration for increasing values of time scale separation $a$ and network size $N$}. Directed random network with $K=32$. Average initial conditions: $\theta(0)=0.09$, $\Gamma(0)=0.75$ with $W(0)=1$, input $I=0.1$. (a) Parameters $a=10^{4}$ and $b=8\times 10^{-2}$; (b) $a= 10^{5}$ and $b=10^{-2}$ and (c) $a= 10^{6}$ and $b= 10^{-3}$. Fitted exponent relation ($m_{\text{fitted}}$) and distance to criticality coefficient
($dcc=|m_{\text{theory}}-m_{\text{fitted}}|$) also shown.}
\label{fig7}
\end{center}
\end{figure*}

The distributions of avalanche sizes and durations for the homeostatic system
are presented in figure~\ref{fig5}.
In figure~\ref{fig6}, we show how the finite-size scaling improves with 
increasing time scale separation $a$. Both the collapse of the curves is enhanced and the 
characteristic bump decreases as $a$ grows.
Moreover, figure~\ref{fig7} shows that the exponent's relation also tends to agree 
with the theory for increasing separation of time scales $a\to\infty$, resulting 
in a small distance to criticality ($dcc<0.01$) for $a = 10^6$. 
These results show that the homeostatic mechanism proposed here is fully capable
of approaching the underlying critical point ($\tau=3/2$, $\tau_d=2$ and $m=2$).
This is different from other dynamics from the literature, since previous studies either
lead to unclear avalanche distributions, or fail to reproduce
the average size and duration scaling law.

The collapse of the avalanche size distribution can be calculated by, first,
rescaling the variable through $u=s/s_c$, where $s_c=q_1N^D$ is the cutoff size of avalanches
and $D$ is the size dimensionality (not to be confused with avalanche duration).
Then, we define the scaling function
$\mathcal{G}(u)=q_2s^\tau P(u)$. If the system is critical, this function will collapse the data from
the simulations of different system sizes by plotting $\mathcal{G}(u)$ vs. $u$.
The same can be achieved for
the duration $d$, and for the cumulative distributions as well.
The quantities $q_1$ and $q_2$ are called the metric coefficients;
$q_1$ is obtained by collapsing the tail of the distributions on top of
one another, and $q_2$ is obtained by fitting of a given distribution
to the simulation data.
These are not universal quantities, changing from one distribution to another~\cite{Pruessner2012}.
For the avalanche exponents (and hence, criticality) to be well-defined,
these coefficients must not depend on the system size $N$~\cite{Pruessner2012}.
In the data shown in figure~\ref{fig6}, the failure of the collapse for small $a$
indicates that the metric coefficients depend on $N$.
However, this happens in a non-trivial way, since the collapse enhances for increasing $a$.
This tells us that the dependence on $N$ also becomes negligible in the growing $a$ limit,
as the system gets closer and closer to the critical point.
This is analytically supported by the form of the fixed points in 
equations~\eqref{rhoFP}--\eqref{sthetafp};  numerically by the $h$ amplitude in figure~\ref{fig2},
as well as by the $dcc$ in figure~\ref{fig7}.

The dependence of the metric coefficients on $N$ for small $a$ is expected
due to the quasi-critical nature of our homeostatic system.
We can explain it by the
superposition of avalanches as follows. First, notice that the
probability of uncorrelated spikes increases with $N$ since every neuron is an independent 
stochastic unit. These random spikes are spontaneously generated by the network
due to a finite external field $h$, and they act as seeds to the avalanches. Therefore,
the separation between events decreases with growing $N$ and, hence, avalanche
superposition increases (see the Supplementary Information for more details).
As our activity measure $\rho(t)$ cannot distinguish
avalanches emerging from different seeds, this leads to an increased number of 
larger avalanches for larger $N$ (for more details, see the Supplementary Material).
This effect vanishes for increasing $a$ because such regime promotes the separation
of the internal time scales of the system. This in turn generates a decreasing
effective external field $h\to0$, ceasing simultaneous seeding of avalanches.

\section{Discussion}

We have presented a neuronal network model that self-organizes toward quasi-criticality,
even in the presence of non-zero inputs $I_i$. This is an important result, given that 
cortical neurons, for example, are constantly bombarded by input from various areas. 
The homeostatic thresholds $\theta_i$ -- which can also be interpreted as adaptation 
currents \cite{Benda2010} --  lead to $|h| < 10^{-4}$, an (almost exact) 
adaptation to the inputs. That is, instead of imposing $h=0$ as is usually done
in standard SOC models~\cite{Dickman2000},
here we construct homeostatic mechanisms such that the effective fields
$h_i(t) = I_i - \theta_i(t)$ tend towards zero.
This is no mere detail, but a crucial ingredient for a truly
self-organized critical model~\cite{Dickman2000}.

The $\tilde{W}^{*}$ component of the fixed point is always subcritical, but tends 
to the critical value when $ab\to\infty$. 
From a biological perspective, staying in the vicinity of a
subcritical state might be advantageous to decrease the risk of spontaneous
runaway activity~\cite{Priesemann2015}. Such activity could be linked
to dysfunctional regimes like epilepsy. 
Also, from the perspective of information processing and task performance, 
it was recently shown that while criticality might be beneficial for complex computational tasks,
it can be detrimental to the performance in simpler ones~\cite{Cramer2020}. 
This suggests that the subcriticality expressed by our model,
with just a few excursions to a critical state for some particular tasks,
might be the best regime for the network.

The form of the adaptation currents adopted for thresholds $\theta_i$ is known 
to generate negative correlations between consecutive interspike 
intervals~\cite{Nesse2021}. This means that long silent intervals will be 
followed by shorter ones, on average. In addition, experimental results suggest 
that large avalanche-like events are followed, on average, by small events 
in the resting state of the brain~\cite{Lombardi2013}.
Thus, the adaptation to the external field displayed by our model
could be the mechanism responsible for the generation
of the separation between time scales. This separation has
to be imposed in all classic SOC models, but emerges naturally
in our model for large $a$.

After the self-organization process, the system hovers around a stable (quasi-critical) 
fixed point with small amplitude orbits, minimizing the large stochastic 
oscillations observed in previous models~\cite{Costa2017,Kinouchi2019,Girardi2020}. 
In particular, the system gets closer to criticality as the timescale ratio 
$a = \tau_\theta/\tau_W= \tau_\theta / \tau_\Gamma$ increases. This effect is related to the
decrease of the effective external field $h$.
In turn, this vanishing $h$ is what restores the true scaling and power laws in the avalanche
distributions as the separation of time scale grows. This is because both $a\to\infty$
and $h\to0$ are required to suppress the superposition of avalanches.

Self-organization mechanisms that tune neuronal networks to various states
based on the homeostatic interplay of external input and intrinsic
activity have already been explored in the literature~\cite{Zierenberg2018}.
However, scale-free avalanches with mean-field DP-like exponents that satisfy the scaling
relation had been only observed in the limit of vanishing external drive~\cite{Decandia2021}.
In our model, both the exponents and the scaling arise naturally from the adaptation
mechanism. 

Regarding the unavoidable fine-tuning $A\approx 1$ imposed in our study, we need to
remember Hernandez-Urbina and Herrmann~\cite{Hernandez2017,Kinouchi2020}:
fine-tuning a hyperparameter in local homeostatic
mechanisms is very different from globally fine-tuning the parameters $\{W_{ij},h_i\}$ in the
original static model. In any case, a challenge to the
community persists: is it possible to obtain $A\approx 1$ without any form of fine-tuning?
We conjecture that this is impossible: the need for $h^* \approx 0$ will impose strict
conditions similar to $A \approx 1$ to any other homeostatic model~\cite{Girardi2021}.

Generalizing our results, it is plausible that any system under the influence of external
fields (say, a magnetic field $H$ in spin systems)
-- as in earthquakes, forest fires, voting or epidemic models based on spins, cellular
automata or even continuous time integrate-and-fire dynamics -- can achieve a near-critical
regime through the inclusion of adaptive local fields 
(as our $\theta_i(t)$) whose timescale should be much slower than that of the rest of
the system. 

Nonetheless, it is not clear how time-varying  inputs $I_i(t)$  would affect the behavior 
of our system. We conjecture that the thresholds $\theta_i(t)$ would  produce a phenomenon
akin to sensory adaptation~\cite{Petermann2009, Hahn2010}. 
If so, for short time scales, our homeostatic networks would respond to the derivatives of
the external signal, as opposed to signal intensity. This could lead to yet unknown
computational properties. 
For example, the results on the optimization of the dynamic range in critical networks~\cite{Kinouchi2006}
could be challenged, or, at least,
would need to be reconsidered in the context of sensory systems with adaptation.
This important issue will be studied in a future extended paper.

 \section{Acknowledgments}
G.M. would like to thank CAPES for financial support.
M.G.-S. thanks the financial support of NSERC grant BCPIR/493076-2017
from A. Longtin and L. Maler.
O.K. acknowledges CNAIPS-USP and CNPq, Conselho Nacional de Desenvolvimento Científico e Tecnológico support.
This work was produced as part of the activity of FAPESP Research, Innovation and Dissemination Center for Neuromathematics (grant \#2013/07699-0 S. Paulo Research Foundation).

\bibliographystyle{unsrt}
\bibliography{ourbib}

\end{document}